\documentclass[namedreferences]{solarphysics}
\usepackage[hyperref,optionalrh,solaromanenum]{spr-sola-addons} % For Solar Physics 
\usepackage{graphicx}                    % For eps figures, newer & more powerfull
\usepackage{color}                       % For color text: \color command
\usepackage{breakurl}                    % For breaking URLs easily trough lines
\usepackage{lscape} %page rotation	
\newcommand{\aap}{{\it Astron. Astrophys.}}
\newcommand{\apj}{{\it Astrophys. J.}}
\newcommand{\apjl}{{\it Astrophys. J. Letter}}
\newcommand{\nat}{{\it Nature}}
\newcommand{\pasj}{{\it PASJ}}
\newcommand{\solphys}{{\it Solar Phys.}}

\begin{document}

\begin{article}

\begin{opening}

\title{Velocity Structure and Temperature Dependence of Extreme-Ultraviolet Jet Observed by Hinode}

%%%%%%%%%%%%%%%%%%%%%%%%%%%%%%%%%%%%%%%%%%%%%%%%%%%
%% Authors Names
%
\author[addressref={aff1},corref,email={t.kawai@isee.nagoya-u.ac.jp}]{\inits{T.}\fnm{T.}\lnm{Kawai}}
\author[addressref={aff1},corref]{\inits{N.}\fnm{N.}\lnm{Kanda}}
\author[addressref={aff1},corref]{\inits{S.}\fnm{S.}\lnm{Imada}}

%%%%%%%%%%%%%%%%%%%%%%%%%%%%%%%%%%%%%%%%%%%%%%%%%%%
%% Runningheads
%
%\runningauthor{}
%\runningtitle{}

%%%%%%%%%%%%%%%%%%%%%%%%%%%%%%%%%%%%%%%%%%%%%%%%%%%
%% Affilations 
%% id should be the same with \author addressref value.
\address[id={aff1}]{Institute for Space-Earth Environmental Research, Nagoya University, Furo-cho, Chikusa-ku, Nagoya, Aichi, Japan}

%%%%%%%%%%%%%%%%%%%%%%%%%%%%%%%%%%%%%%%%%%%%%%%%%%%
%%% Abstract 
\begin{abstract}
The acceleration mechanism of EUV/X-ray jets is still unclear. 
For the most part, there are two candidates for the mechanism. 
One is magnetic reconnection, and the other is chromospheric evaporation. 
We observed a relatively compact X-ray jet that occurred between 10:50 -- 11:10 UT on February 18, 2011 by using the Solar Dynamics Observatory/Atmospheric Imaging Assembly, and the X-ray Telescope, Solar Optical Telescope, and EUV Imaging Spectrometer aboard Hinode. 
Our results are as follows:
1) The EUV and X-ray observations show the general characteristics of X-ray jets, such as an arch structure straddling a polarity inversion line, a jet bright point shown at one side of the arch leg, and a spire above the arch. 
2) The multi-wavelength observations and Ca II H-band image show the existence of a low-temperature ($\sim$~10 000K) plasma (i.e., filament) at the center of the jet. 
3) In the magnetogram and Ca II H-band image, the filament exists over the polarity inversion line and arch structure is also straddling it. In addition, magnetic cancellation occurs around the jet a few hours before and after the jet is observed. 
4) The temperature distribution of the accelerated plasma, which was estimated from Doppler velocity maps, the calculated differential emission measure, and synthetic spectra show that there is no clear dependence between the plasma velocity and its temperature. 
For the third result above, observational results suggest that magnetic cancellation is probably related to the occurrence of the jet and filament formation. 
This result suggests that the trigger of the jet is magnetic cancellation rather than an emerging magnetic arch flux. 
The fourth result indicates that acceleration of the plasma accompanied by an X-ray jet seems to be caused by magnetic reconnection rather than chromospheric evaporation. 
\end{abstract}

%%%%%%%%%%%%%%%%%%%%%%%%%%%%%%%%%%%%%%%%%%%%%%%%%%%
%% Keywords
%

\end{opening}
%-------------------------------------------------

%%%%%%%%%%%%%%%%%%%%%%%%%%%%%%%%%%%%%%%%%%%%%%%%%%%
%% Sections
%
\section{Introduction}
	\label{s:introduction} 

Jet-shaped eruptions are often observed in extreme-ultraviolet (EUV) regions and/or in soft X-rays. 
These eruptions are called EUV/X-ray jets. 
The jets are characterized by nearly collimated high-speed flows.
Many studies have shown that EUV jets and X-ray jets are associated with each other.
The fundamental physical process of EUV and X-ray jets is believed to be the same.
The first observations of X-ray jets \citep[e.g.,][]{Shibata92,Strong92} were carried out with the soft X-ray telescope (SXT; \cite{Tsuneta91}) on board Yohkoh (\cite{Ogawara91}). 
X-ray jets consist mainly of hot ($>$ a few MK) and fast ($\sim$ a few 100 km s$^{-1}$) flows, and magnetic reconnection is believed to be one of the drivers for the jet.
An early physical model of the jet was presented by \cite{Shibata94}.
In their model, the magnetic reconnection which occurs between the emerging flux and a pre-existing coronal field causes the jet. 
Typically, two types of flows are expected in this model. 
One is the flow directly accelerated by magnetic reconnection, and the other flow is related to chromospheric evaporation. 
The flow-related magnetic reconnection is accelerated by a magnetic tension force, and its velocity almost reaches the Alfven velocity.
On the other hand, the flow related to chromospheric evaporation is accelerated by a pressure gradient force, and its theoretical upper limit is approximately 2.35 $C_s $, where $C_s$ is the sound speed \citep[][]{Fisher84}. 
So far, these flows and associated jets were well reproduced by numerical simulations \citep[e.g.,][]{Yokoyama95}.
One of the observable differences between the reconnection flow and chromospheric evaporation is the temperature dependence.
It is generally believed that magnetic reconnection flows are independent of temperature, although chromospheric evaporation is a temperature-dependent flow.
In a large flare, several studies have confirmed both chromospheric evaporation \citep[e.g.,][]{Milligan09, Imada15} and magnetic reconnection flows \citep[e.g.,][]{Innes03, Imada13}, individually.
By contrast, in the case of small jets, it is generally difficult to distinguish which flow signature is associated with magnetic reconnection or chromospheric evaporation in the jet structure. 
Temperature dependence of the flows has been discussed to distinguish the origin of the flow (e.g., \cite{Imada07}, \cite{Imada11b}, \cite{Tian12}). 
\cite{Matsui12} attempted to reveal the acceleration mechanism of jets by using multi-wavelength spectroscopic observations; they claimed that the acceleration is caused by chromospheric evaporation and magnetic reconnection simultaneously.

Standard jets are primarily discussed in two dimensions.
\cite{Moore10} discussed polar X-ray jets in three dimensions and found that there are two types of the jets, one is a standard jet and the another is called ``blowout jet'' which needs a three-dimensional structure. 
In their model, the arch structure of magnetic fields straddles the polarity inversion line, the core field also straddles the inversion line inside of the arch, and open fields exist around them, before jet outbreaks. 
Magnetic fields that are strongly sheared emerge inside of the arch structure. 
Then, the arch is pushed up, and the current sheet is formed between the arch and the open field. 
After that, magnetic reconnection occurs, and the eruption begins. 
At one side of the leg of the arch, which forms the current sheet, small arches of heated magneto fields are built; this is called jet bright point (JBP). 
JBPs are observed clearly in the X-ray band. 
Reconnections between the arch and open fields stretch the core arch and reconnection site, which form a wide (curtain-like) spire, and continue to form a JBP. 
Then, a tether-cutting reconnection occurs between the core arch and the sheared emerging flux, which builds a flare arcade that straddles the polarity inversion line. 
Finally, the emerging flux and the open fields are reconnected, which also forms arch structures and causes filament eruption. 
After the jet develops, the arch, the open field, and JBP, which are reconnection-heated, are observable. 
Only blowout jets have sheared and twisted magnetic fields and are accompanied by filament eruptions; these are the main differences between standard jets and blowout jets.
From coronal EUV and X-ray observations, \cite{Moore13} indicated that the number of standard jets and blowout jets are almost the same. 

Recently another model of jets was discussed by \cite{Sterling15}. 
Before the jet occurs, there is an arch structure that straddles the polarity inversion line, a twisted magnetic field which winds small-scale filaments (mini-filaments) side by side, and open fields also exist. 
In this model, the jet is a result of the mini-filament eruption. 
This eruption is accompanied by a small flare, which is seen as the JBP, beneath the erupting mini-filament. 
Because of this, the location of the JBP is different from the location predicted by the \cite{Moore10} picture. 
The cause of the mini-filament eruption is magnetic field cancellation, emerging flux, or something else, while the \cite{Moore10} idea is based on the emerging flux model.
Subsequently, reconnections occur between twisted fields and open fields, which build a heated arch structure and spire. 
Such a mini-filament eruption model was numerically reproduced by \cite{Wyper18} using three-dimensional magnetohydrodynamic simulations. 
\cite{Sterling15} also said that both standard jets and blowout jets are basically caused by same mechanism, mini-filament eruption. 
When mini-filament eruption is largely confined inside of the jet-base region, it results in a standard jet. 
On the other hand, when mini-filament eruption is ejective, it results in a blowout jet. 
Moreover, there are two apparent differences between \cite{Moore10} and \cite{Sterling15}. 
One is that filament eruption occurs inside of the arch in \cite{Moore10} but occurs outside of the arch in \cite{Sterling15}. 
Another is the direction of the spire; it approaches the JBP in \cite{Moore10} but moves away from the JBP in \cite{Sterling15}. 

\cite{Kamio07} derived velocity structures of north-polar jets in coronal holes by spectroscopic observations. As a result, blue-shifted plasma is located above the bright loops and red-shifted plasma is located at the foot of the bright loops.
\cite{Madjarska11} found that a temperature of 12 MK and density of $4 \times 10^{10} \mathrm{cm}^{-3}$ in the energy deposition region of the X-ray jet occurred in the quiet Sun. These results are obtained from spectroscopic observations of Fe XXIII (263.76 \AA) and a pair of Fe XII lines. 
\cite{Young15} found jet-like structures, which are named dark jets, in Doppler velocity maps obtained by spectroscopic observations in coronal holes that cannot be seen by imaging instruments. Although they claimed that these dark jets are as common as regular coronal-hole jets, the mass flux seemed to be approximately two orders of magnitude lower.

In this paper, we discuss the acceleration mechanism of plasma by an X-ray jet that occurred with magnetic cancellation and formation of a filament. 
To do this, we estimate the temperature distribution of the accelerated plasma by using Doppler velocity maps obtained from spectroscopic observations and differential emission measures of the jet estimated from multi-wavelength EUV observations. 

\section{Data and observations}

\begin{figure} 
\centerline{\includegraphics[width=0.75\textwidth,clip=]{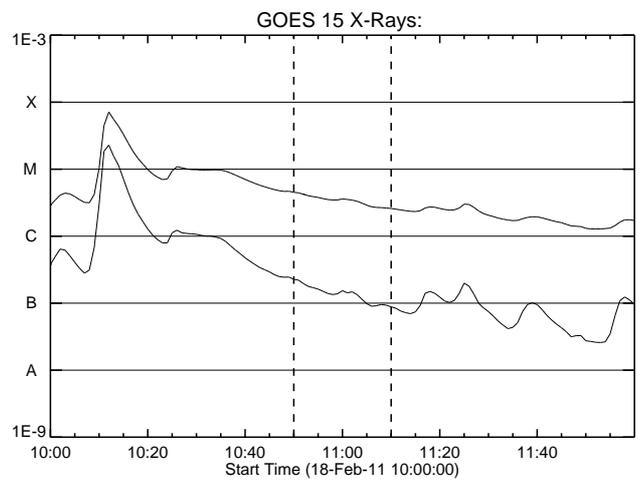}}
\caption{Time evolution of X-rays obtained from GOES 15. Vertical dashed lines represent when the jet occurs and disappears.}\label{fig:GOES}
\end{figure}

We observed a jet that occurred between 10:50 -- 11:10 UT on February 18, 2011 in NOAA active region 11158. 
The jet was relatively compact (50'' $\times$ 50''), and it continued for about 20 min. 
The temporal variation of X-rays obtained from Geostationary Operational Environmental Satellite (GOES) 15 is shown in Figure~\ref{fig:GOES}.
The vertical dashed lines in the figure represent the time interval of the jet. 
There are small enhancements in X-rays around 11:00 UT, which is associated with the EUV/X-ray jet. 
The jet is located near the southwest limb $(X, Y)\sim(750'', -250'')$, and plasma associated with the jet erupts toward the north. 
The spire of the jet orients nearly parallel to the surface rather than standing up vertically. 
This means that the ambient coronal fields along which the spire develops are leaning sideways unlike usually depicted in cartoons or simulation results. 
However, the fundamental dynamics of the jet should be same because the ambient fields constraining the spire are still in the corona even though it is not standing vertically.

\subsection{Data}
	\label{sub:data}
	
We used Solar Dynamics Observatory (SDO)/Atmospheric Imaging Assembly (AIA;~\cite{Lemen12}) to determine the entire time evolution of the jet.
The SDO/AIA instrument acquired full-Sun images with a spatial resolution of ~1000 km. 
We used the AIA spectral channels at 94, 131, 171, 193, 211, 304, 335, and 1600 \AA~to analyze the high temporal evolution of the jet ($\sim$24 s in 1600 \AA~and $\sim$12 s in other bands). 
Each of the AIA channels has different temperature coverage. 
The AIA 304, 171, 193, 211, 335, 94, 131, and 1600-\AA~channels represent 10$^{4.7}$, 10$^{5.8}$, 10$^{7.3}$, 10$^{6.3}$, 10$^{6.4}$, 10$^{6.8}$, 10$^{7.0}$, 10$^{4.0}$ K plasma in the case of flares (\cite{boe}; \cite{Lemen12}), respectively. 

The X-Ray Telescope (XRT \citep{Golub07}) on board Hinode, which has a high spatial resolution ($\sim$2 arcsec) and a wide temperature coverage ($6.1 < \log{T} < 7.5$), was also used to determine the time evolution of the jet. 
The XRT observed the jet in each 156-s interval with a Thin Be filter, which has a maximum temperature response of approximately 10 MK \citep[][]{Narukage14}.

AIA and XRT data were processed by the \verb|aia_prep| and \verb|xrt_prep| routines in SolarSoftWare (SSW,~\cite{Freeland98}) to eliminate instrumental effects caused by warm, hot, and dusty pixels, the dark current pedestal, and others.

The EUV Imaging Spectrometer (EIS) aboard Hinode is a high spectral- and spatial-resolution spectrometer aimed at studying dynamics in the corona \citep{Culhane07}.
The Hinode EIS observed the EUV jet in a slit-scanning mode with a 2" wide slit and exposure duration of 5 s at each scanning point. 
During the jet, EIS successfully obtained EUV images and line-of-sight (LOS) velocities (Doppler shift) in several emission lines with $\sim$1500 km spatial resolution. 
By using a series of Fe emission lines (from Fe X ($Te \sim 10^{5.8}$) to Fe XV ($Te \sim 10^{6.3}$)), we can reveal the temperature dependence of the fine-scaled structure and dynamics associated with the jet. 
EIS data from the raster were processed using the EIS software provided by the team, which corrects for flat field, dark current, cosmic rays, hot pixels, and slit tilt.
For thermal reasons, there was an orbital variation of the line position, causing an artificial Doppler shift of $\pm$20 km s$^{-1}$ that has a sinusoidal behavior.
This orbital variation of the line position was corrected using the housekeeping data \citep[][]{Kamio10}.

\subsection{AIA observations}
	\label{sub:AIA}

\begin{figure} 
\centerline{\includegraphics[width=1.0\textwidth,clip=]{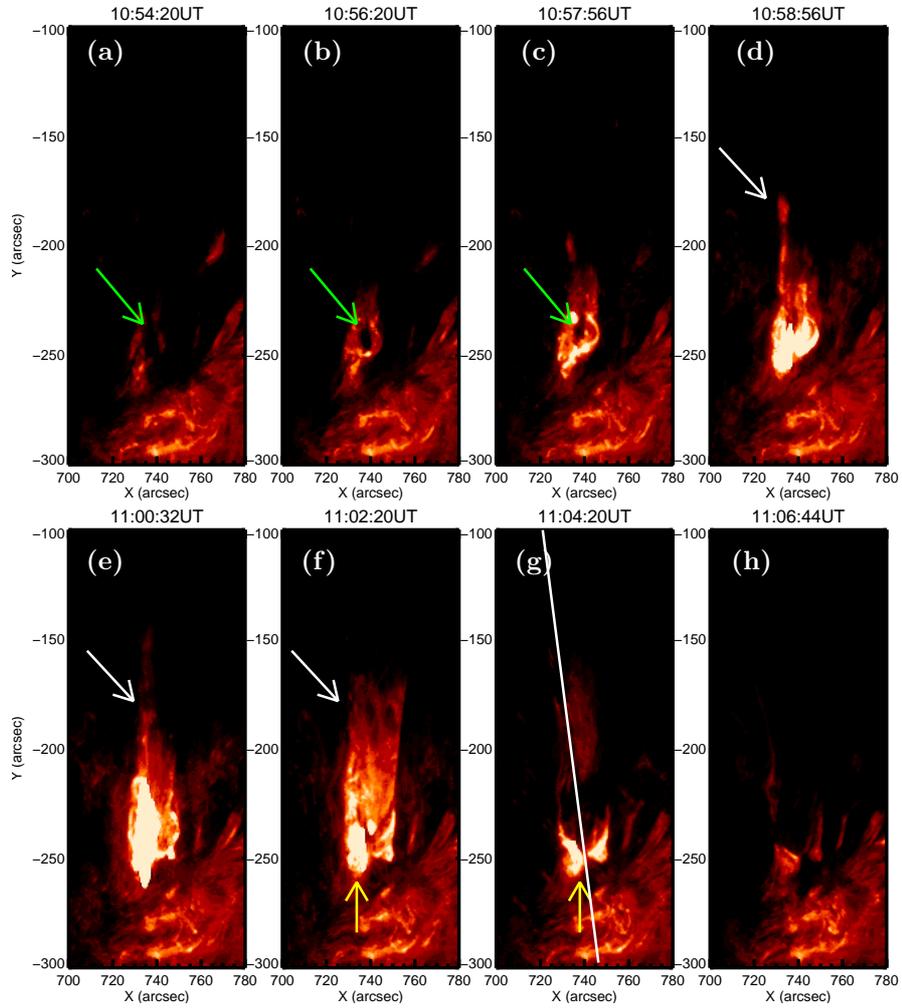}}
\vspace{-1.06\textwidth}    
     \centerline{\bf    
      \hspace{0.068 \textwidth}  \color{white}{(a)}
      \hspace{0.169\textwidth}  \color{white}{(b)}
      \hspace{0.169\textwidth}  \color{white}{(c)}
      \hspace{0.169\textwidth}  \color{white}{(d)}
         \hfill}
\vspace{0.52\textwidth}
     \centerline{\bf       
      \hspace{0.068 \textwidth}  \color{white}{(e)}
      \hspace{0.169\textwidth}  \color{white}{(f)}
      \hspace{0.169\textwidth}  \color{white}{(g)}
      \hspace{0.169\textwidth}  \color{white}{(h)}
         \hfill}
\vspace{0.50\textwidth}
\caption{Temporal evolution of the jet obtained by SDO/AIA at 304 \AA. Dark region surrounded by bright region is observed from (a) to (c) (green arrows). Spire is clearly observable in (d) and (e) (white arrows). Yellow arrows point to a jet bright point (JBP). White line in (g) approximately indicates the jet direction.}
\label{fig:AIA304}
\end{figure}

In order to study the temporal evolution of the EUV jet, we analyzed AIA data of February 18, 2011 from 10:50 to 11:10 UT. 
The development of the EUV jet can be clearly seen in AIA 304 \AA~movie (movie S1).
Figure~\ref{fig:AIA304} shows snapshots of the temporal evolution of the jet from immediately before the occurrence to after the decay observed by AIA at 304 \AA. 
The jet started around 10:54 (Figure~\ref{fig:AIA304} (a)) and finished around 11:06 (Figure~\ref{fig:AIA304} (g)). 
The typical structures of blowout jets, such as JBP (Figure~\ref{fig:AIA304} (b)), spire (Figure~\ref{fig:AIA304} (d)), and development into a wide curtain-like spire (Figure~\ref{fig:AIA304} (f)), were clearly observed. 
Based on the model of a blowout jet \citep[][]{Moore10}, there were magnetic fields that had an arch structure straddling a dark region equivalent to the sheared filament (Figure~\ref{fig:AIA304} from (a) to (c): green arrows). 
Firstly, the current sheet formed between ambient magnetic fields and the arch structure fields, and magnetic reconnection occurred there. 
As the magnetic reconnection developed, magnetic fields were heated and a spire formed (Figure~\ref{fig:AIA304} (d): white arrows). 
Then, the sheared filament reconnected with the outer magnetic fields and erupted to the Solar-Y (northward) direction. 
A curtain-like spire and JBP also formed at the same time owing to the reconnection with outer magnetic fields (Figure~\ref{fig:AIA304} (f): yellow arrows).

\begin{figure} 
\centerline{\includegraphics[width=1.0\textwidth,clip=]{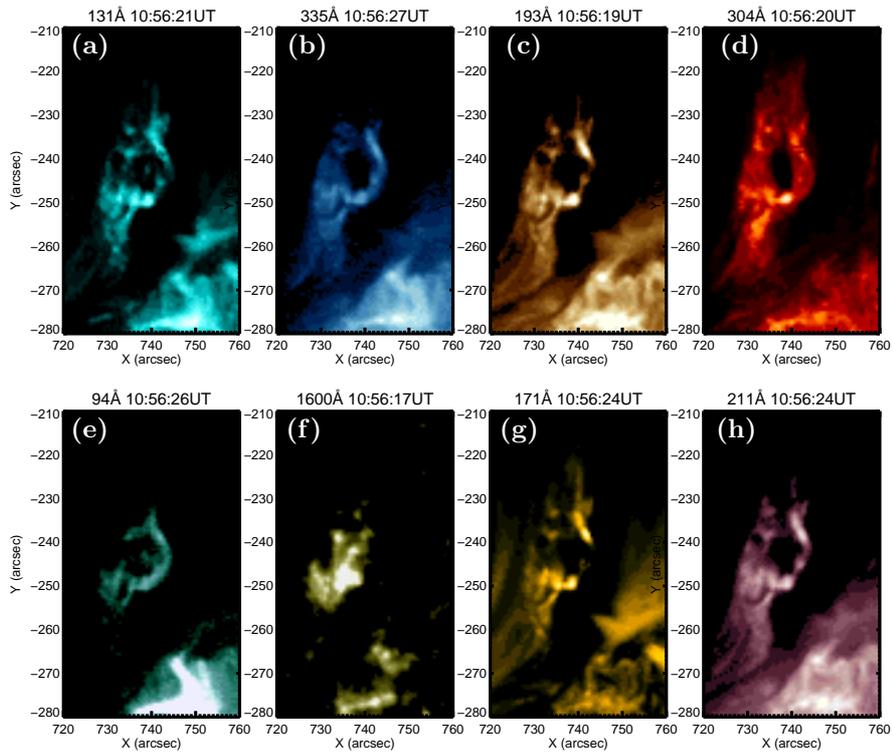}}
\vspace{-0.80\textwidth}    
     \centerline{\bf       
      \hspace{0.057 \textwidth}  \color{white}{(a)}
      \hspace{0.169\textwidth}  \color{white}{(b)}
      \hspace{0.169\textwidth}  \color{white}{(c)}
      \hspace{0.169\textwidth}  \color{white}{(d)}
         \hfill}
\vspace{0.382\textwidth}
     \centerline{\bf       
      \hspace{0.057 \textwidth}  \color{white}{(e)}
      \hspace{0.169\textwidth}  \color{white}{(f)}
      \hspace{0.169\textwidth}  \color{white}{(g)}
      \hspace{0.169\textwidth}  \color{white}{(h)}
         \hfill}
\vspace{0.50\textwidth}
\caption{Result obtained by AIA in multiple wavelengths. Upper row, observations at 131, 335, 193 and 304 \AA. Lower row, observations at 94, 1600, 171 and 211 \AA.}\label{fig:MULTI_AIA}
\end{figure}

The AIA images in the different bands show the temperature difference in the temporal evolution of the EUV jet.
Figure~\ref{fig:MULTI_AIA} shows a snapshot of the jet with multi-wavelength bands between 10:56:17 and 10:56:27 UT (Figure~\ref{fig:MULTI_AIA}), corresponding to the beginning of the jet, which expanded around the dark region. 
From the top-left in Figure~\ref{fig:MULTI_AIA}, 131, 335, 193, 304, 94, 1600, 171, and 211 \AA~images are shown. 
We can observe jet-like structures, similar to 304 \AA~, in all images except for 1600 \AA.
In all the bands except for 1600\AA~(f), there are dark void structures surrounded by a brighter jet. 
On the contrary, in the image of 1600\AA, a bright region can be seen where dark regions typically exist in other bands.

\begin{figure} 
\centerline{\includegraphics[width=1.0\textwidth,clip=]{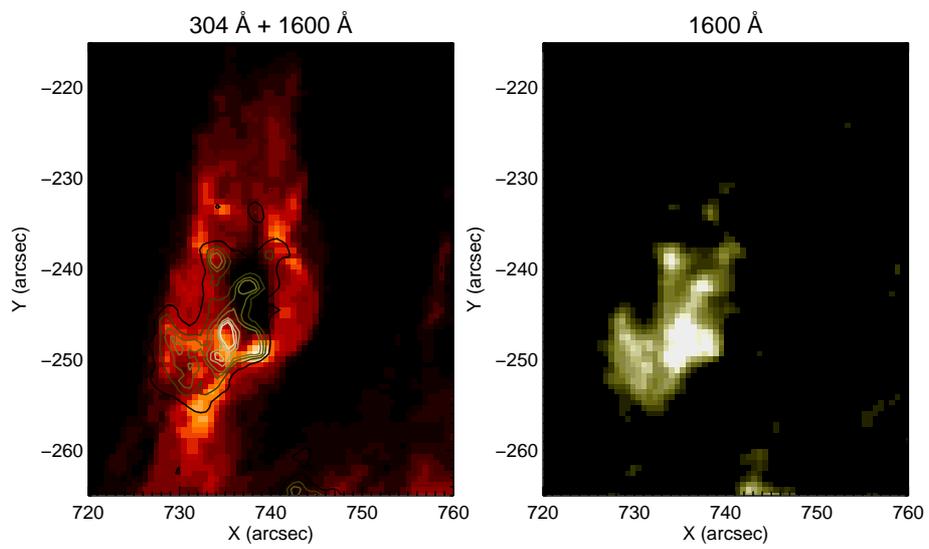}}
\caption{Left panel: Depicts contours of observations at 1600 \AA~over the image obtained at 304 \AA. Location of bright region observed at 1600 \AA~corresponding to a dark region observed at 304 \AA. Right panel: The image obtained at 1600 \AA.}
\label{fig:over}
\end{figure}

The left panel of Figure~\ref{fig:over} shows an enlarged image to highlight the jet from Figure~\ref{fig:MULTI_AIA} (d), overlaid with contours of intensity of the image of Figure~\ref{fig:MULTI_AIA} (f).
The right panel also shows an enlarged image of Figure~\ref{fig:MULTI_AIA} (f).
Although there is a three-second gap between the two images, the bright region in 1600 \AA~ clearly exists in place of the dark region in 304 \AA~.
Therefore, these images imply that there is a cool ($\sim$~10 000 K) plasma, i.e., a filament, in the center of the EUV jet.

\begin{figure} 
\centerline{\includegraphics[width=1.0\textwidth,clip=]{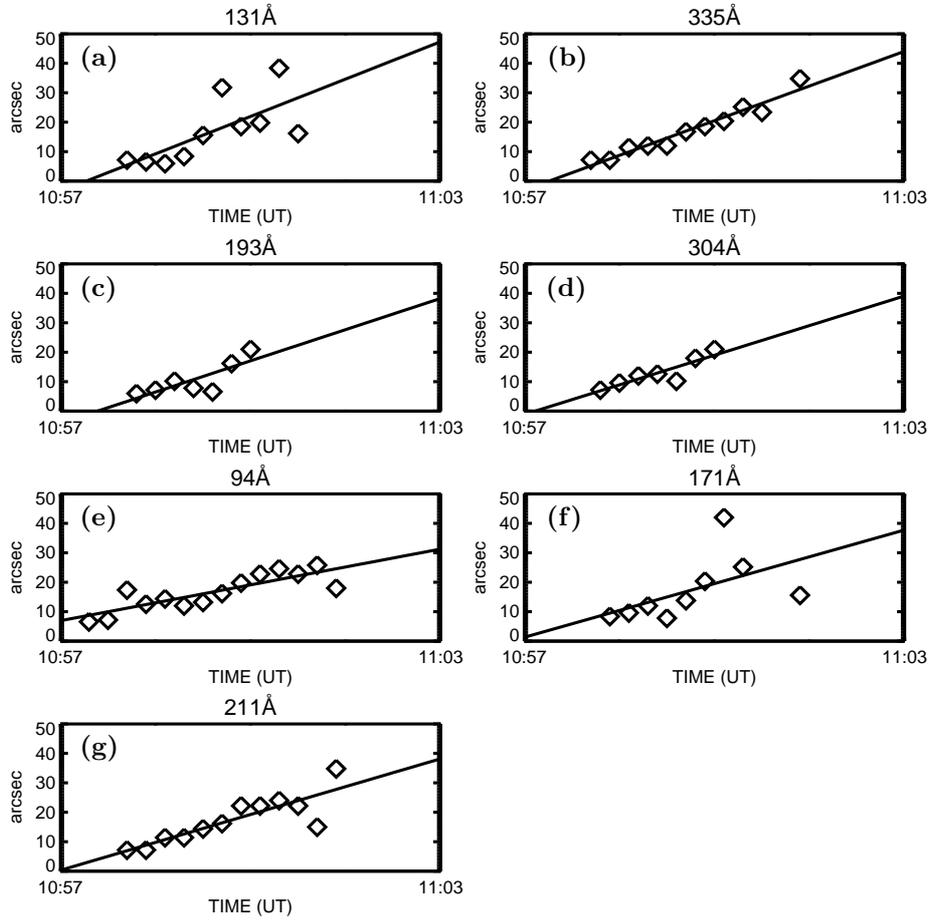}}
\vspace{-0.96 \textwidth}    
     \centerline{\bf       
      \hspace{0.07 \textwidth}  \color{black}{(a)}
      \hspace{0.44\textwidth}  \color{black}{(b)}
         \hfill}
\vspace{0.215\textwidth}    
     \centerline{\bf       
      \hspace{0.07\textwidth}  \color{black}{(c)}
      \hspace{0.44\textwidth}  \color{black}{(d)}
         \hfill}
\vspace{0.215\textwidth}
     \centerline{\bf    
      \hspace{0.07\textwidth}  \color{black}{(e)}
      \hspace{0.44\textwidth}  \color{black}{(f)}
         \hfill}
\vspace{0.215\textwidth}    
     \centerline{\bf       
      \hspace{0.07\textwidth}  \color{black}{(g)}
         \hfill}
\vspace{0.19\textwidth}
\caption{Time-distance plot along white line in Figure~\ref{fig:AIA304} (g). Diamond symbols represent the distance between upper border of bright region and jet base. Solid lines are fitted by least squares method, with which we calculated vertical line-of-sight velocities.}
\vspace{0.1\textwidth}
\label{fig:vel_multi}
\end{figure}

To estimate the apparent velocity of the jet, we created a time--distance plot along the white line in Figure~\ref{fig:AIA304} (g). 
This white line approximately indicates the jet direction. 
It is defined by the vertical bisector between both geometric centers of the bright regions seen in the decay phase (e.g., Figure~\ref{fig:AIA304} (g)). 
The horizontal axes of Figure~\ref{fig:vel_multi} represent UT during the occurrence of the jet and the vertical axes represent distance along the white line of Figure~\ref{fig:AIA304}(g). 
Diamond symbols represent the distances between the upper border of the bright region and the base of the jet along the white line of Figure~\ref{fig:AIA304}(g).
Sometimes diamond symbols are not plotted because we removed the data obtained when the telescope was in flare-mode.
Each apparent velocity is calculated from the incline of each solid line, which is fitted to the diamond symbols by the least-squares method.
Table~\ref{tab:aia} shows the results of the apparent velocities in each wavelength.
The apparent velocity of the jet in 1600 \AA~ cannot be calculated because the arch legs of Figure~\ref{fig:AIA304}(g) do not exist.

\subsection{XRT observations}
	\label{sub:XRT}

\begin{figure} 
\centerline{\includegraphics[width=1.0\textwidth,clip=]{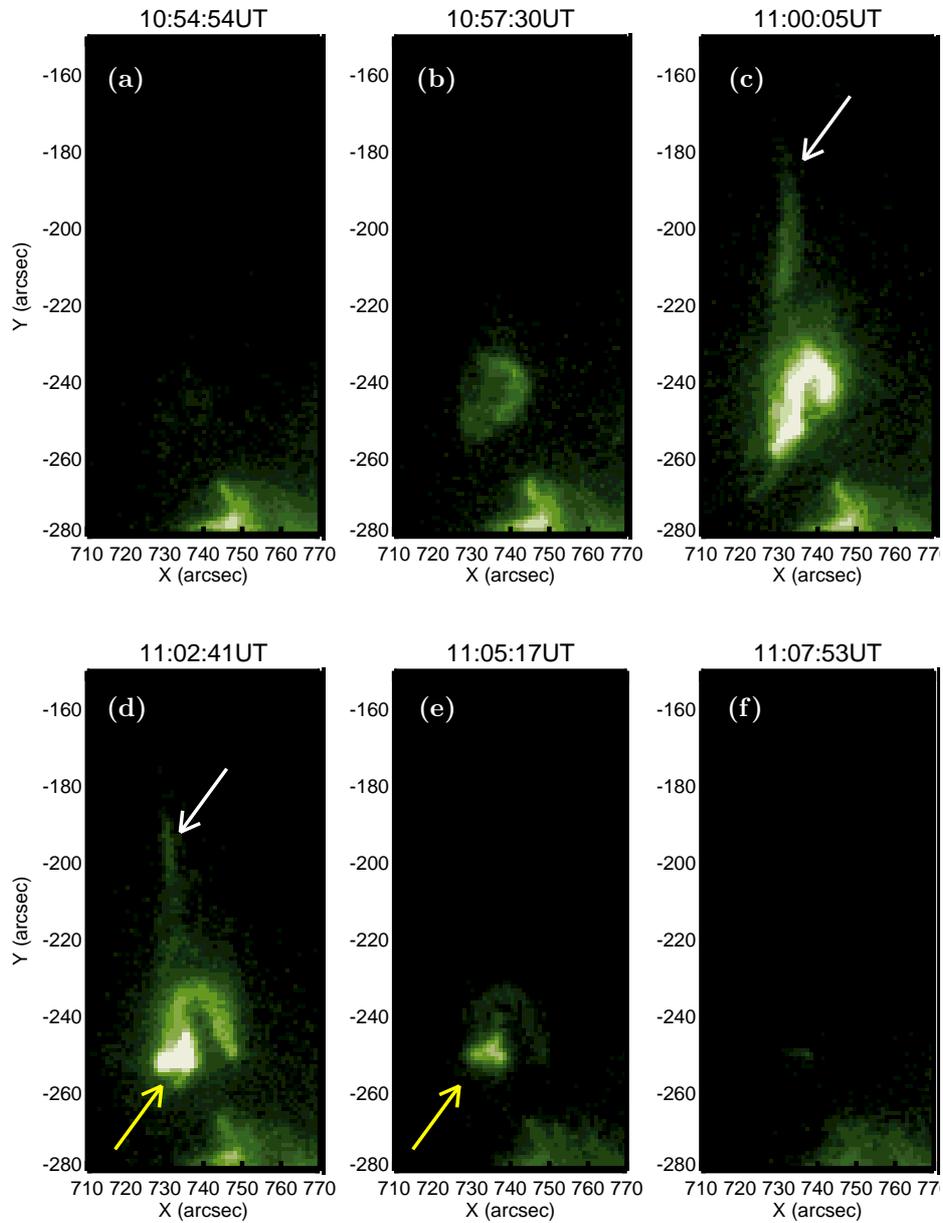}}
\vspace{-1.3\textwidth}    
     \centerline{\bf    
      \hspace{0.09 \textwidth}  \color{white}{(a)}
      \hspace{0.27\textwidth}  \color{white}{(b)}
      \hspace{0.27\textwidth}  \color{white}{(c)}
         \hfill}
\vspace{0.65\textwidth}    
     \centerline{\bf    
      \hspace{0.09 \textwidth}  \color{white}{(d)}
      \hspace{0.27\textwidth}  \color{white}{(e)}
      \hspace{0.27\textwidth}  \color{white}{(f)}
         \hfill}
\vspace{0.59\textwidth}
\caption{Time evolution obtained from Hinode/XRT with Thin Be filter. White and yellow arrows represent spire and JBP, respectively. JBP still exists after jet decays (e).}\label{fig:XRT}
\end{figure}

Figure~\ref{fig:XRT} shows the results obtained from the observations by the XRT with the Thin Be filter. 
The maximum temperature response of this filter is at approximately 10 MK \citep[][]{Narukage14}.
Figures 6 (a) and (b) show the solar corona before the EUV jet was observed in soft X-rays. 
In Figures 6 (c) and (d), there is an arch structure, and a spire (white arrows) extends along to Solar-Y (northward) direction. 
In Figure 6 (d), a relatively bright region, which represents a JBP (yellow arrows) is observed at the left base of the arch. 
The spire disappeared and the arch structure decayed in Figure~\ref{fig:XRT} (e); however, the JBP still appeared bright, which implies that the JBP is at a much higher temperature than other regions. 

Unlike the images obtained by AIA, the curtain-like structure of the spire cannot be seen (Figure~\ref{fig:AIA304} (f) and Figure~\ref{fig:XRT} (d)) and the arch structure can be seen more clearly (Figure~\ref{fig:AIA304} (e) and Figure~\ref{fig:XRT} (c)). 
Moreover, the JBP is much brighter than the other leg, which is more obvious than in the AIA images.

\subsection{EIS observations}
	\label{sub:EIS}
	
\begin{figure} 
\centerline{\includegraphics[width=1.0\textwidth,clip=]{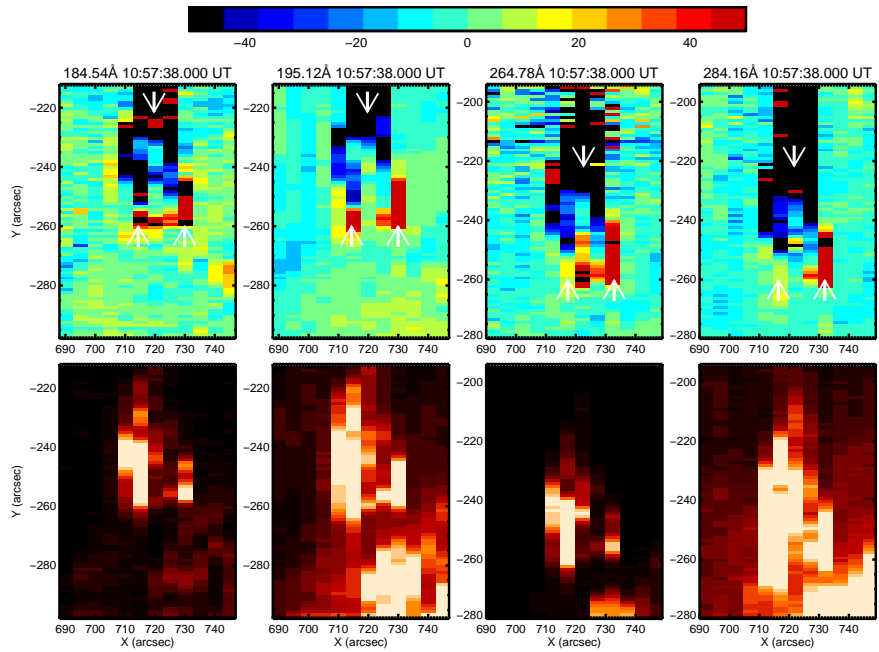}}
\caption{Upper row shows Doppler velocity maps and lower row shows intensity maps around the jet in Fe X (184.54 \AA), Fe XII (195.12 \AA), Fe XIV (264.78 \AA) and Fe XV (284.16 \AA) obtained from Hinode/EIS. Both of the legs of the arch have a redshift and the spire has a blueshift. We obtained each line profile at the region indicated by white arrow.}\label{fig:eis_all}
\end{figure}

Figure~\ref{fig:eis_all} shows the Doppler velocity maps and intensity maps around the jet observed by Hinode/EIS with 184.54\AA~(Fe X;~$\log_{10}T \sim 6.00$),~ 195.12\AA~(Fe XII;~$\log_{10}T \sim 6.10$),~ 264.78\AA~(Fe XIV;~$\log_{10}T \sim 6.30$) and~ 284.16\AA~(Fe XV;~$\log_{10}T \sim 6.30$)~\citep{Culhane07} from 10:57:38 UT to 11:03:25 UT. 
Matching the result of AIA, flows of plasma along the open magnetic field (spire) form blueshifts, and reverse flows create redshifts, at the both ends of the arch. 
There is a gap in the range of the observations along Solar-Y (north-south) direction between the short wavelength side (184.54\AA~and 195.12\AA) and the long wavelength side (264.78\AA~and 284.16\AA.) 

\begin{figure} 
\centerline{\includegraphics[width=1.0\textwidth,clip=]{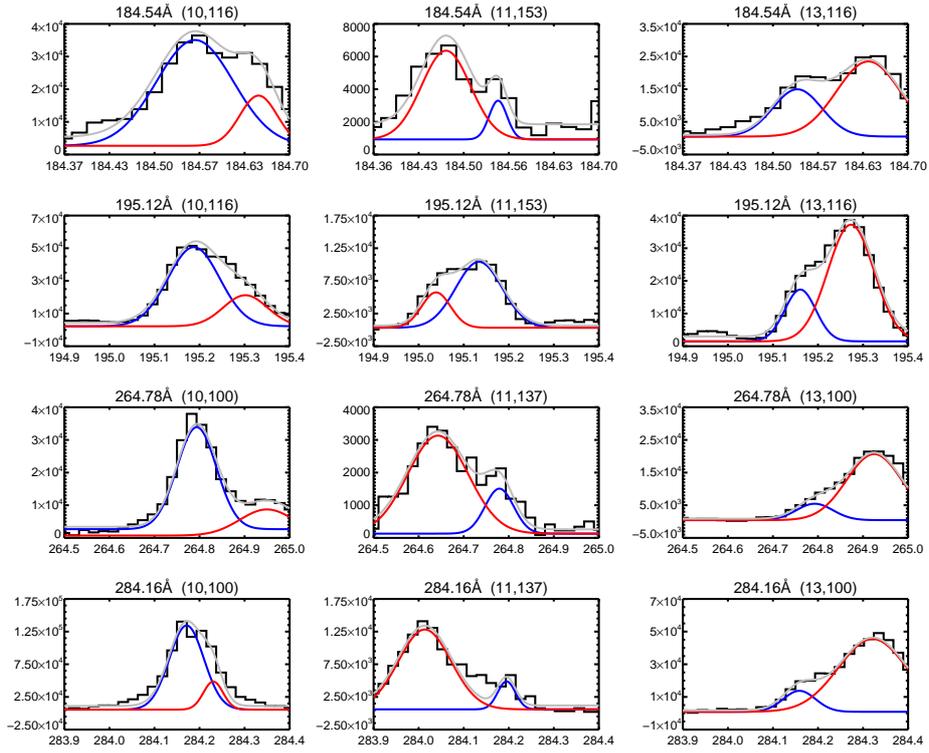}}
\caption{Line profile at the region of Figure~\ref{fig:eis_all} indicated by white arrow. Left, center, and right column represent those of the JBP, spire, and right leg of the arch, respectively. Each column represents each wavelength window. Profiles are fitted with double Gaussians. Each gray line represents sum of red and blue lines.}\label{fig:line}
\end{figure}

Figure~\ref{fig:line} shows line profiles in the region, where each white arrow of Figure~\ref{fig:eis_all} indicates each wavelength window. 
The columns on the left, center, and right represent those of the JBP, spire, and right leg of the arch, respectively. 
Each row corresponds to each wavelength window. 
These profiles are fitted by double Gaussian curves to separate the profiles into plasma with a high and low Doppler velocity (red and blue lines). 
Each gray line represents the sum of the red and blue lines.
Plasma with Doppler blueshift can be seen precisely at the same location as the spire (center column), and it can also be seen with Doppler redshift where both of the legs of the jet appeared (left and right columns) in all wavelength windows. 
Table~\ref{tab:vel} shows Doppler velocities for each region and wavelength. 
A positive value represents the redshift direction.

To study how plasma is accelerated in the jet, we derived temperature distributions of radiative intensity at each wavelength emitted from the accelerated plasma, as follows:
\begin{enumerate}
	\item Estimate the differential emission measure (DEM) of the jet from six wavelength observations of AIA (131\AA, 335\AA, 193\AA, 94\AA, 171\AA~, and 211\AA) by using the code developed by \cite{Aschwanden11} for each region (spire, left leg, and right leg). 
	The calculated temperature range and its resolution are from 5.7 to 6.3, and 0.05 in log scale, respectively. 
	\item Derive the synthetic spectra radiated from the accelerated plasma in each region and in each bin of temperature (regarding DEM outside the bin as zero) from the estimated DEM by using the atomic database CHIANTI and \verb|ch_synthetic.pro| (\cite{DelZanna15}). 
	The wavelength resolution is set at 0.02\AA.
	The electron density is assumed to be the root square of the emission measure, which is the DEM integrated over temperature within the calculated range.
	The default CHIANTI ionization equilibrium table is used.
	Only the radiation values from Fe X, Fe XII, Fe XIV, and Fe XV are calculated. 
	\item Extract the intensities whose wavelengths are 184.54\AA, 195.12\AA, 264.78\AA, and 284.16\AA~ from the derived spectra for each temperature bin.
	\item Calculate the rate of which temperature band of plasma emitted the radiation at each wavelength.
\end{enumerate}

\begin{figure} 
\centerline{\includegraphics[width=0.75\textwidth,clip=]{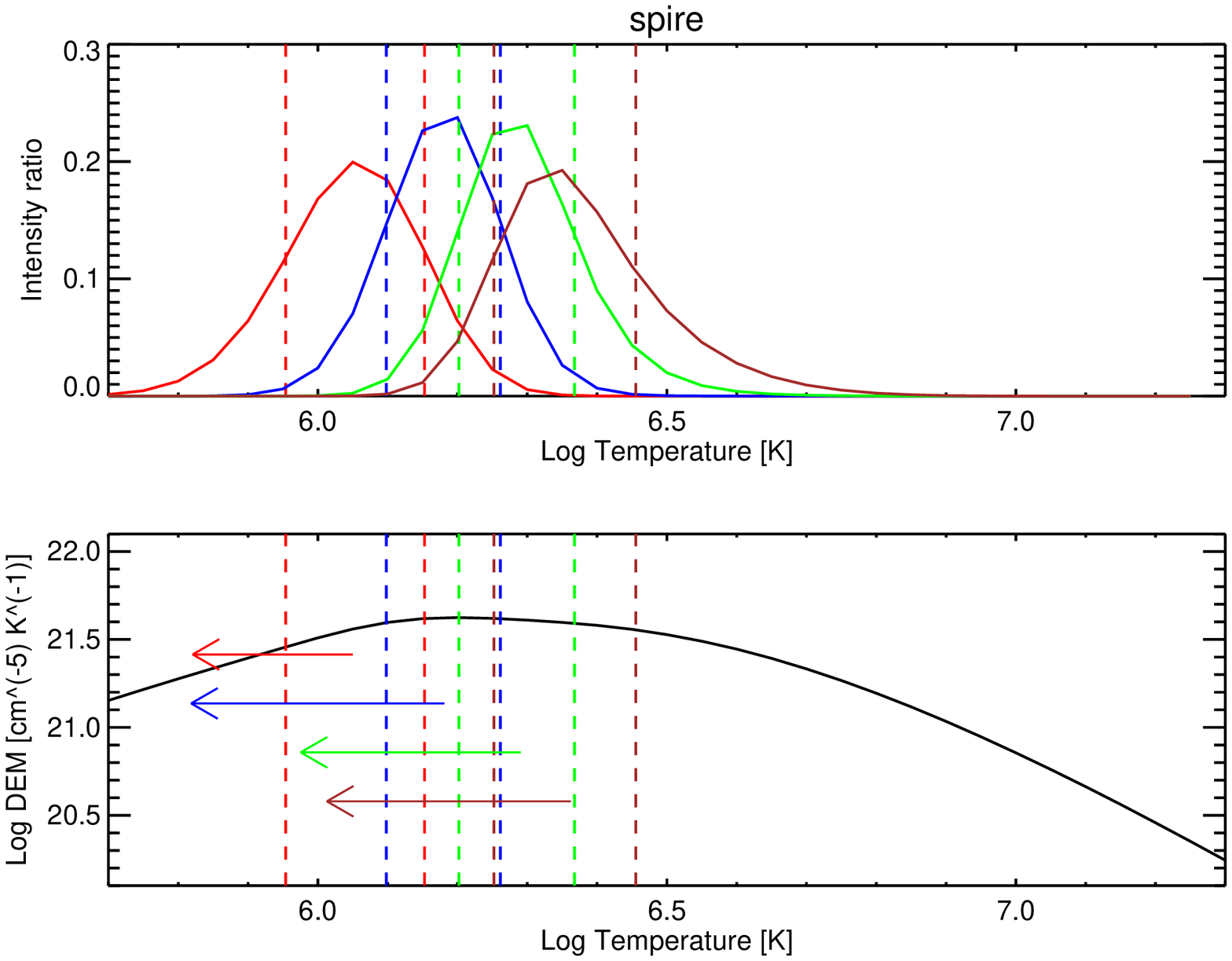}}
\centerline{\includegraphics[width=0.75\textwidth,clip=]{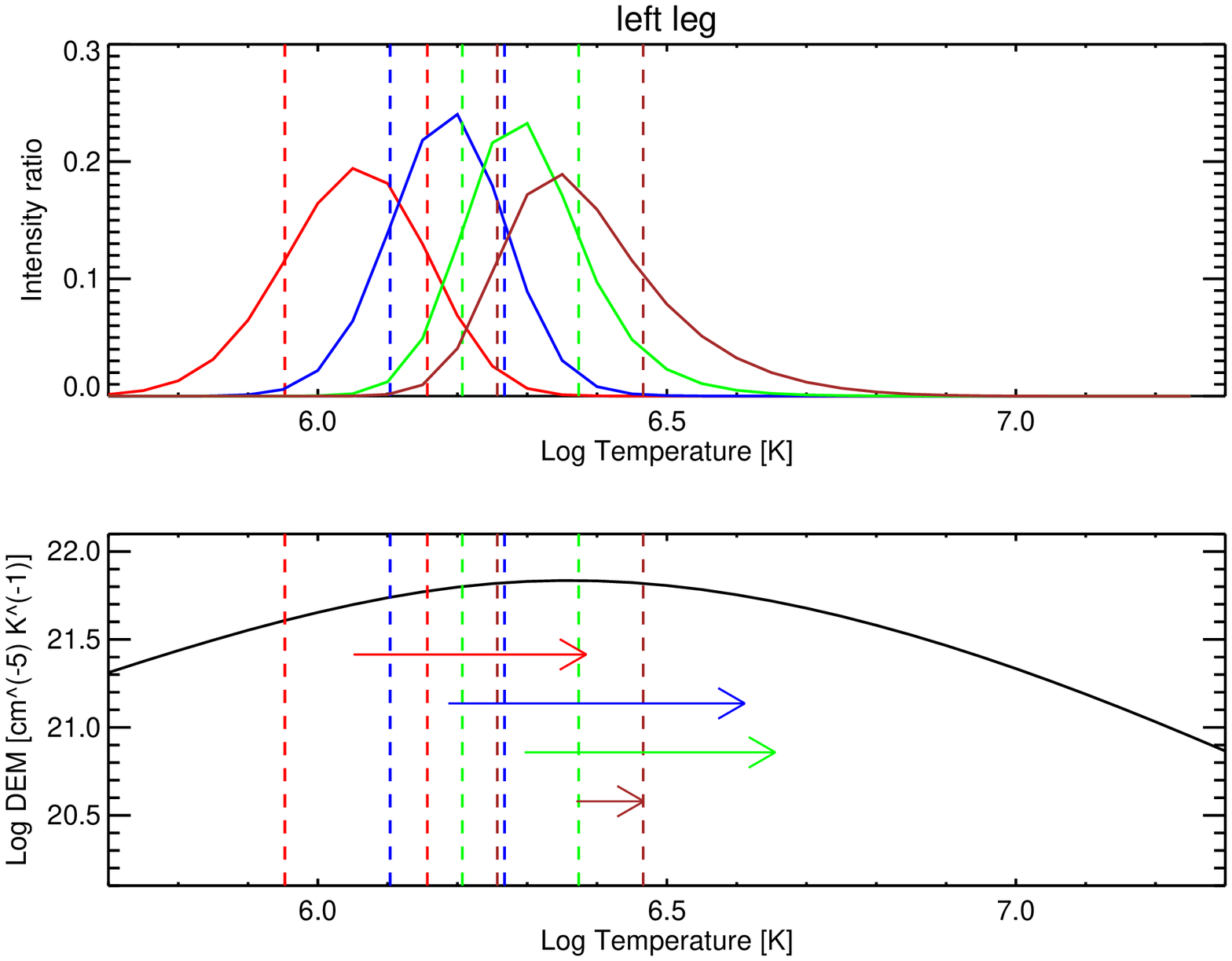}}
\centerline{\includegraphics[width=0.75\textwidth,clip=]{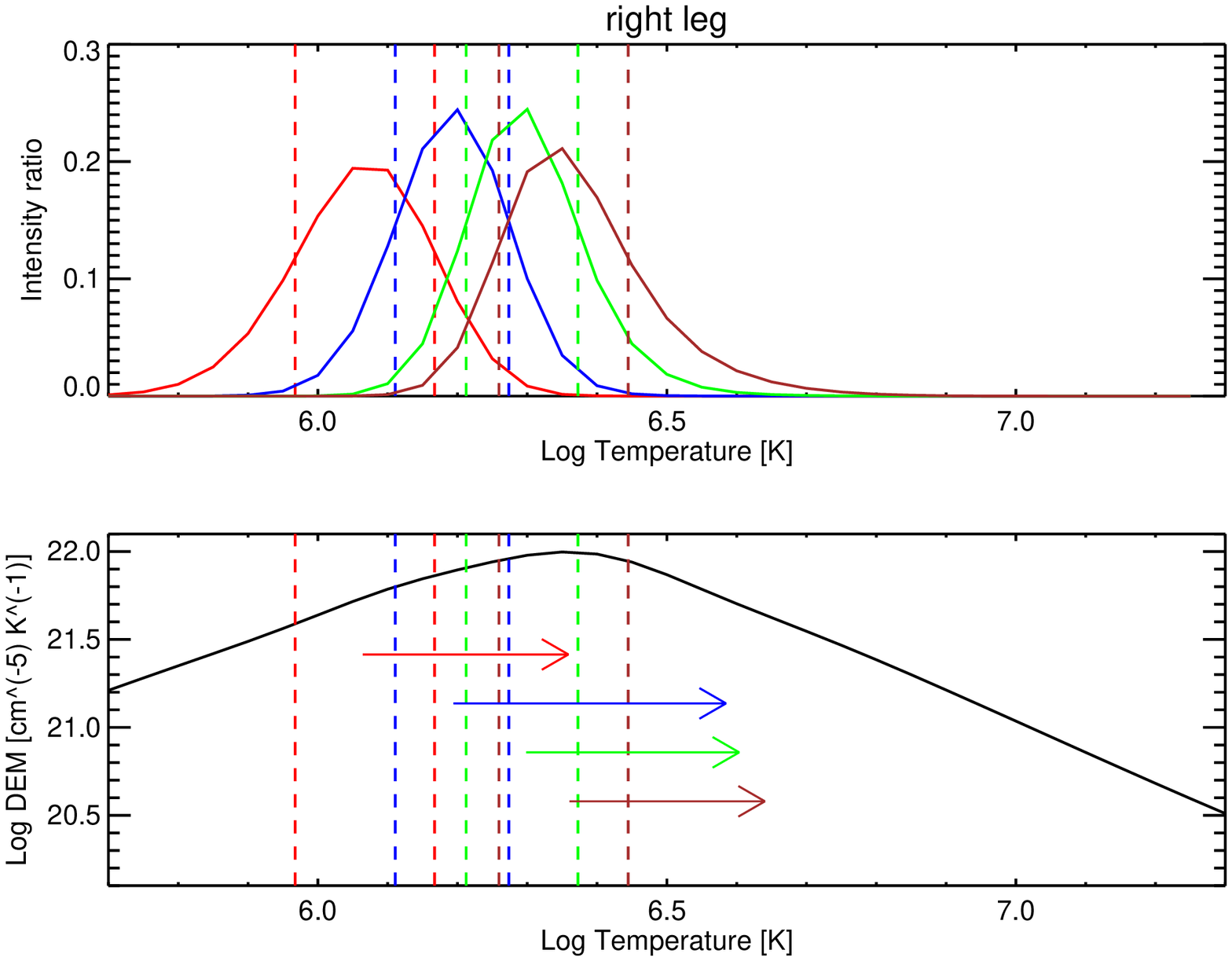}}
\vspace{-1.67\textwidth}  
     \centerline{\bf   
      \hspace{0.78 \textwidth}  \color{black}{(a)}
         \hfill}
\vspace{0.25\textwidth}    
     \centerline{\bf   
      \hspace{0.78\textwidth}  \color{black}{(b)}
         \hfill}
\vspace{0.25\textwidth}    
     \centerline{\bf 
      \hspace{0.78\textwidth}  \color{black}{(c)}
         \hfill}
\vspace{0.25\textwidth}    
     \centerline{\bf   
      \hspace{0.78 \textwidth}  \color{black}{(d)}
         \hfill}
\vspace{0.25\textwidth}    
     \centerline{\bf   
      \hspace{0.78\textwidth}  \color{black}{(e)}
         \hfill}
\vspace{0.25\textwidth}    
     \centerline{\bf   
      \hspace{0.78\textwidth}  \color{black}{(f)}
         \hfill}
\vspace{0.2\textwidth}
\caption{Temperature distributions of plasmas which radiate each wavelength emission (a, c, e) and DEM (b, d, f) in spire (a, b), left leg (c, d) and right leg (e, f), respectively. Red, blue, green and brown lines indicate 184.54\AA~(Fe X), 195.12\AA~(Fe XII), 264.78\AA~(Fe XIV) and 284.16\AA~(Fe XV), respectively. Each dashed line represents $1 \sigma$ of Gaussian fitting of each temperature distribution. Arrows represent Doppler velocity measured by EIS. Left- and right-pointing arrows represent blue shift and red shift, respectively.}\label{fig:pops}
\end{figure}

According to the above, we obtained plasma temperature distributions of radiative intensity at each wavelength. 
Figures~\ref{fig:pops} (a), (c), and (e) show the temperature distributions of radiative intensities estimated from the DEM and the spectra in the spire, left leg, and right leg, respectively. 
Red, blue, green, and brown lines indicate 184.54\AA~(Fe X), 195.12\AA~(Fe XII), 264.78\AA~(Fe XIV), and 284.16\AA~(Fe XV), respectively. 
Each solid line in (b), (d), and (f) indicates the DEM calculated by the code of \cite{Aschwanden11} using AIA data, and each arrow represents the Doppler velocity obtained from EIS (a right-pointing arrow represents plasma accelerated toward the downward direction). 
Each range between a pair of dashed lines represents $\pm{1\sigma}$ of each temperature distribution, where $\sigma$ is the standard deviation of the temperature distribution fitted by a Gaussian. 
It can be seen that the DEM of the largest part of the accelerated plasma observed by EIS does not change considerably depending on the temperature at each wavelength. 
We found there is no dependence of Doppler velocity on plasma temperature.

\subsection{SOT observations}
	\label{sub:SOT}

\begin{figure} 
\centerline{\includegraphics[width=1.0\textwidth,clip=]{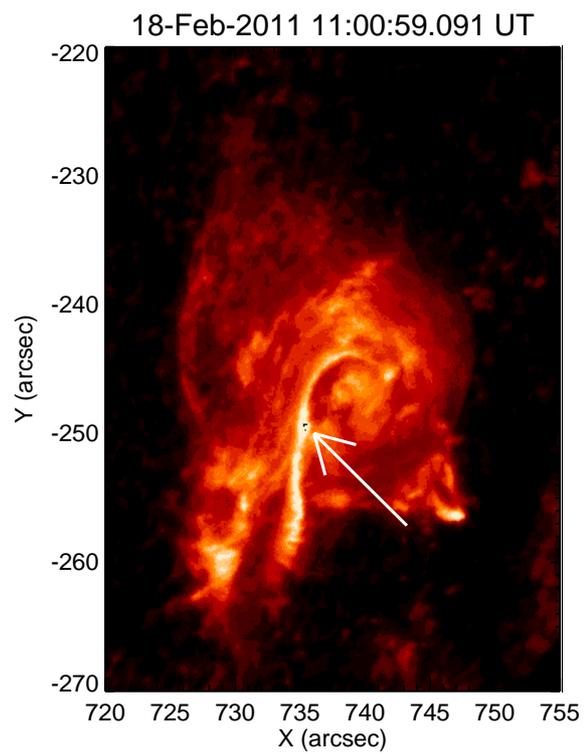}}
\caption{Result of observation obtained from Hinode/SOT with Ca II H band. There is a strongly bright curve, which is thought to be a filament (white arrow).}\label{fig:Ca2H}
\end{figure}

Here, Figure~\ref{fig:Ca2H} shows an observation image of the chromosphere obtained from the Ca II H band by Hinode/SOT at 11:00:59 UT. There is a bright, bold, curved line from $x=735~\mathrm{arcsec}$, $y=-260~\mathrm{arcsec}$ to $x=740~\mathrm{arcsec}$, $y=-245~\mathrm{arcsec}$, which represents a flux rope with a helical structure. 
This flux rope matches the image obtained from AIA at 1600 \AA~ and pierces the dark region observed with AIA, except at 1600 \AA.

\begin{landscape}
\begin{figure} 
\centerline{\includegraphics[width=1.75\textwidth,clip=]{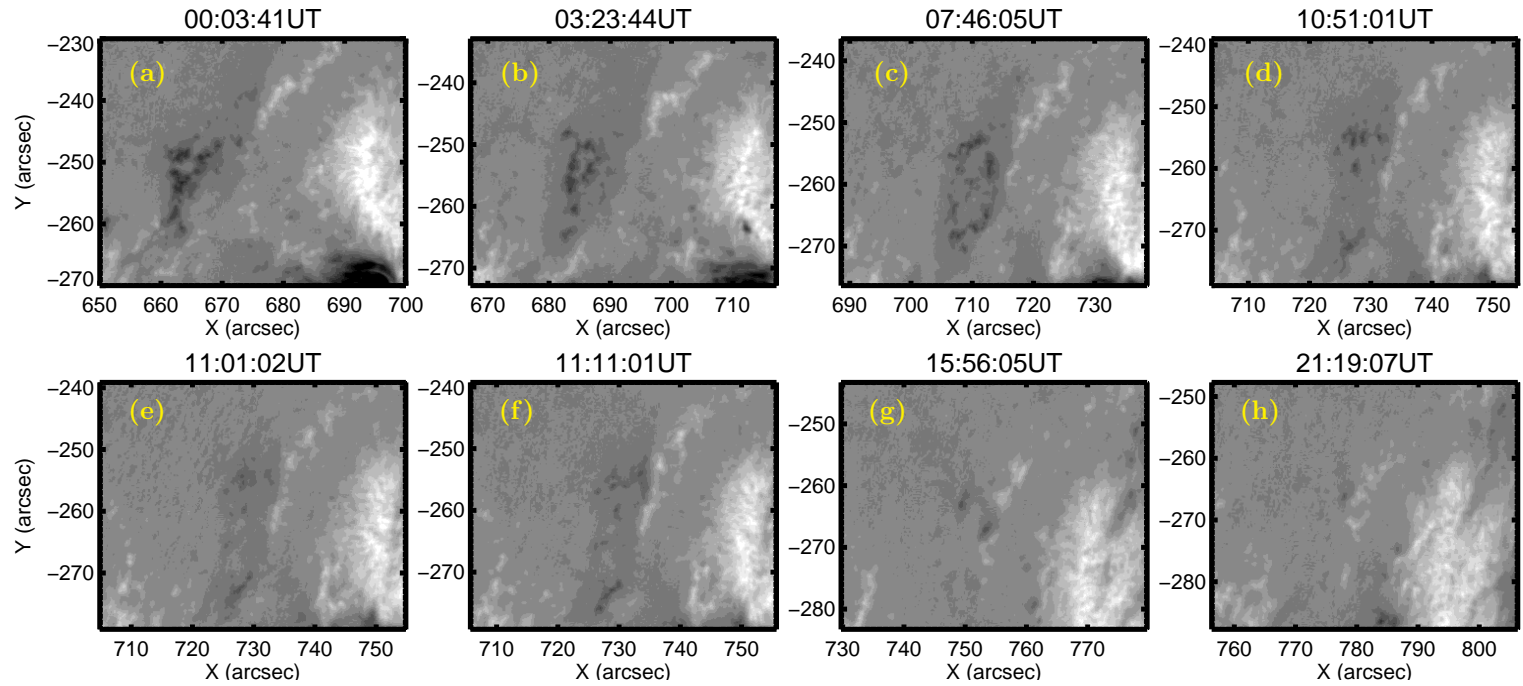}}
\vspace{-0.68\textwidth}    
     \centerline{\bf       
      \hspace{0.062 \textwidth}  \color{yellow}{(a)}
      \hspace{0.34\textwidth}  \color{yellow}{(b)}
      \hspace{0.34\textwidth}  \color{yellow}{(c)}
      \hspace{0.34\textwidth}  \color{yellow}{(d)}
         \hfill}
\vspace{0.33\textwidth}
     \centerline{\bf       
      \hspace{0.062 \textwidth}  \color{yellow}{(e)}
      \hspace{0.34\textwidth}  \color{yellow}{(f)}
      \hspace{0.34\textwidth}  \color{yellow}{(g)}
      \hspace{0.34\textwidth}  \color{yellow}{(h)}
         \hfill}
\vspace{0.33\textwidth}
\caption{Time evolution of magnetic conditions obtained from Hinode/SOT. White regions represent positive flux and black regions represent negative flux. From (a) to (c), both white and black regions where jet occurs move closer to each other, and from (d) to (h), black region gradually disappears with time.}\label{fig:mg}
\end{figure}
\end{landscape}

\begin{figure} 
\centerline{\includegraphics[width=0.85\textwidth,clip=]{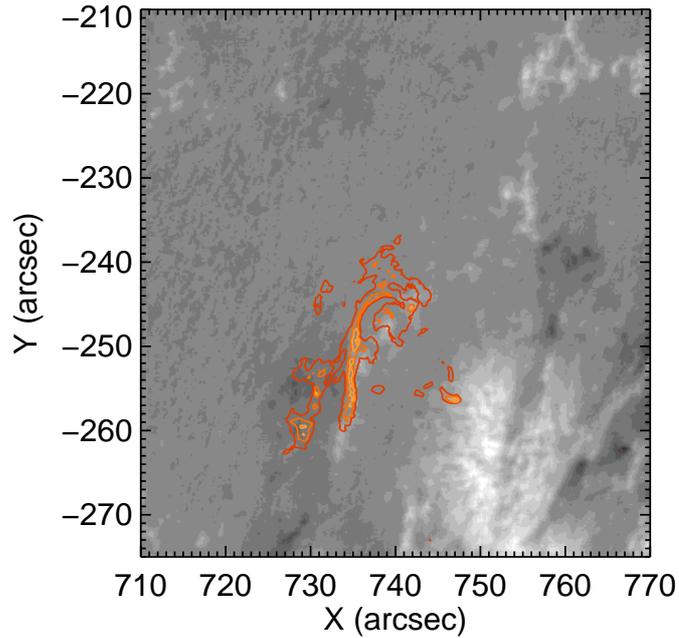}}
\caption{Contours of results obtained in Ca II H band drawn with red and orange lines over the magnetic conditions. Arch structure is seen to straddle polarity inversion line and strongly sheared filament.}\label{fig:mg_fg}
\end{figure}

Figure~\ref{fig:mg} shows the time evolution of the condition of the magnetic fields around the location of the jet. 
White regions represent positive flux and black regions represent negative flux. 
Panels (a), (b), and (c) show the conditions a few hours before the occurrence of the jet.
The jet is occurring from Panels (d) to (f). 
Panels (g) and (h) show conditions a few hours after the jet. 
Figure~\ref{fig:mg_fg} shows the magnetic field map when the jet occurs, as in Figure~\ref{fig:mg}, and intensity contours of the chromospheric image of Figure~\ref{fig:Ca2H} are overlapped with red and orange lines.
It can be seen that the arch structure observed with the Ca II H band is straddling the polarity inversion line.

\begin{figure} 
\centerline{\includegraphics[width=1.0\textwidth,clip=]{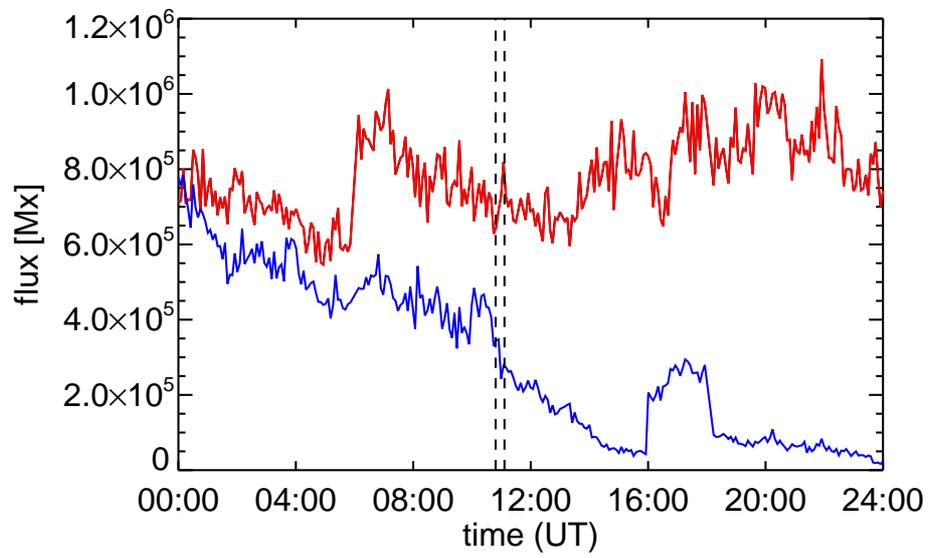}}
\caption{Time evolution of magnetic unsigned flux during the day the jet occurs. Red line, blue line, and dashed lines represents positive flux, negative flux, and the time jet begins and ends, respectively. Negative flux decreases before and after jet occurs.}\label{fig:flux}
\end{figure}

In Figure~\ref{fig:mg}, negative flux regions seem to get smaller with time. 
As a result, we measured the evolution of the unsigned magnetic flux where the jet occurs and around the same region for the whole day the jet occurred. 
To remove the effects of other unrelated elements, we observed the unsigned flux over 30 G. 
Figure~\ref{fig:flux} represents the result. The red, blue, and dashed lines represent positive flux, negative flux, and the times when the jet began and disappeared, respectively. 
According to the result, both negative and positive fluxes definitely decrease before and after the jet occurred, which indicates magnetic cancellation occurred. 
The arch structure and filament observed with the Ca II H band straddles the polarity inversion line in this area, which suggests that magnetic cancellation is related to formation of the filament and occurrence of the jet. 
Positive magnetic fields often increase because sunspots close to them generate positive fields. 
Furthermore, at 16:00 UT, the negative flux suddenly increased because a sunspot came into the measured area.
Apparently, they are unrelated to the trigger of the jet. 

\section{Discussion and summary}
	\label{s:discussion}
We have shown the results of observations of the relatively compact X-ray jet that occurred at 10:50 -- 11:10 UT on February 18, 2011. The data were obtained from SDO/AIA, Hinode/XRT, SOT, and EIS. 
Our results are as follows.

1) EUV and X-ray observations by AIA and XRT show the general characteristics of X-ray jets, such as JBP, spire, arch structure, and the spire expanding to the right and left. 
2) Low-temperature ($\sim$~10 000K) plasma (i.e., a filament) was observed with multi-wavelength observations by AIA and the Ca II H band image by SOT at the center of the jet. 
3) The filament above the polarity inversion line and the arch structure straddling it were found in the magnetogram and the Ca II band image obtained by SOT. Moreover, magnetic cancellation occurred around the jet a few hours before and after the occurrence of the jet. 
4) The velocity of accelerated plasma did not depend on its temperature, according to the temperature distribution calculated from DEM and CHIANTI. 

Comparing observations by AIA, SOT, and XRT to the model of a jet depicted by \cite{Moore10} and \cite{Sterling15}, there are many common features, e.g., an arch structure straddling the polarity inversion line, a JBP positioned at one side of the arch legs, an expanding spire, and an erupted filament. 
It seems to be impossible to recognize the difference between these two models (whether filament eruption occurs inside of the arch or outside, and the direction of expansion of the spire) from observational data. 
Consequently, in this study, both models are applicable. 
As mentioned in point 3), observational results obtained from SOT suggest that magnetic cancellation is probably related to the occurrence of the jet and filament formation. 
In some previous studies (e.g., \cite{Young14a}, \cite{Shen17}, \cite{Sterling17}, \cite{Panesar18}, and \cite{Kumar19}), the magnetic cancelation is observed around the polarity inversion line associated with the jet although it is not observed in all jets.
This result suggests that the trigger of the jet is not the emerging magnetic arch flux as described in \cite{Moore10}; the jet is rather caused by magnetic cancellation as described by \cite{Sterling15}. 

The acceleration mechanism of a coronal jet is still controversial. 
The temperature distribution of the accelerated plasma estimated from Doppler velocity maps obtained from EIS, the DEM calculated by the code developed by \cite{Aschwanden11} with the six-wavelength observations of AIA, and the synthetic spectra derived from CHIANTI code \citep[][]{DelZanna15} together show that there are no clear dependencies between plasma velocity and its temperature. 
Moreover, apparent velocity obtained from AIA multi-wavelength observation has no dependence on plasma temperature.
These results indicate that the acceleration of plasma accompanied by an X-ray jet seems to be caused by magnetic reconnection rather than chromospheric evaporation. 

In this study, we neglected the non-equilibrium ionization effect. 
Generally, plasma is far from equilibrium when plasma is rapidly heated \citep[][]{Imada11a}. 
The violation of the ionization equilibrium naturally causes the increasing of the line formation temperature. 
Non-equilibrium ionization effect in the EUV jet is the important future work.

%%%%%%%%%%%%%%%%%%%%%%%%%%%%%%%%%%%%%%%%%%%%%%%%%%%%%%%%%%%%%%%%%%%%%%%%%%%
%% Appendix
%
% \appendix   

%%%%%%%%%%%%%%%%%%%%%%%%%%%%%%%%%%%%%%%%%%%%%%%%%%%%%%%%%%%%%%%%%%%%%%%%%%%
%% Acknowledgements
%
\begin{acks}[Acknowledgements]
This work was partially supported by the Grant-in-Aid for 17K14401 and 15H05816, and the Program for Leading Graduate Schools, ``PhD Professional: Gateway to Success in Frontier Asia'' by the Ministry of Education, Culture, Sports, Science and Technology. 
Hinode is a Japanese mission developed and launched by ISAS/JAXA, collaborating with NAOJ as a domestic partner, NASA and STFC (UK) as international partners. 
Scientific operation of the Hinode mission is conducted by the Hinode science team organized at ISAS/JAXA. 
This team mainly consists of scientists from institutes in the partner countries. 
Support for the post-launch operation is provided by JAXA and NAOJ (Japan), STFC (U.K.), NASA (U.S.A.), ESA, and NSC (Norway). 
The Solar Dynamics Observatory is part of NASA's Living with a Star program. 
A part of this study was carried by using the computational resource of the Center for Integrated Data Science, Institute for Space-Earth Environmental Research, Nagoya University.
\end{acks}

%%% %%%%%%%%%%%%%%%%%%%%%%%%%%%%%%%%%%%%%%%%%%%%%%%%%%%%%%%%%%%
%% Bibliography
%
% Using BibTeX
%
% \bibliographystyle{spr-mp-sola}
% \bibliography{<bib file>}  

\begin{thebibliography}{99}
 \bibitem[\protect\citeauthoryear{Author}{Year}]{key}
%   <bibliographical entry>
   \bibitem[\protect\citeauthoryear{Aschwanden et al.}{2011}]{Aschwanden11}  
 Aschwanden, M. J., Boerner, P., Schrijver, C. J., \& Malanushenko, A.: 
 2011, \solphys , 283, 5
  \bibitem[\protect\citeauthoryear{Ballegooijen \& Martens}{1989}]{Ballegooijen89}  
 Ballegooijen, A. A. V., \& Martens, P. C. H.: 
 1989, \apj , 343, 984
  \bibitem[\protect\citeauthoryear{Boerner et~al.}{2012}]{boe}  
Boerner, P., et~al.: 
 2012, \solphys , 275, 41
  \bibitem[\protect\citeauthoryear{Culhane et al.}{2007}]{Culhane07}  
 Culhane, J. L., Harra, L. K., James, A. M., et al: 
 2007, \solphys , 243, 61
   \bibitem[\protect\citeauthoryear{Del Zanna et al.}{2015}]{DelZanna15}  
 Del Zanna, G., Dere, K. P., Young, P. R., Landi, E., \& Mason, H. E.:
 2015, \aap, 582, A56
  \bibitem[\protect\citeauthoryear{Fisher et al.}{1984}]{Fisher84}  
 Fisher, G. H., Canfield, R. C., \& McClymont, A, N.: 
 1984, \apj , 281, L79
  \bibitem[\protect\citeauthoryear{Freeland \& Handy}{1998}]{Freeland98}  
 Freeland, S. L., \& Handy, B. N. : 
 1998, \solphys , 182, 500
   \bibitem[\protect\citeauthoryear{Golub et al.}{2007}]{Golub07}  
 Golub, L., Deluca, E., Austin, G., et al.: 
 2007, \solphys , 243, 63
 \bibitem[\protect\citeauthoryear{Imada et al.}{2007}]{Imada07}  
 Imada, S., Hara, H., Watanabe, T., et al.: 
 2007, \pasj , 59, 793
 \bibitem[\protect\citeauthoryear{Imada et al.}{2011a}]{Imada11a}  
 Imada, S., Murakami, I., Watanabe, T., et al.: 
 2011, \apj , 742, 11
 \bibitem[\protect\citeauthoryear{Imada et al.}{2011b}]{Imada11b}  
 Imada, S., Hara, H., Watanabe, T., et al.: 
 2011, \apj , 743, 57
 \bibitem[\protect\citeauthoryear{Imada et al.}{2013}]{Imada13}  
 Imada, S., Aoki, K., Hara, H., Watanabe, T., Harra, L. K., \& Shimizu, T.: 
 2013, \apjl , 776, L11
 \bibitem[\protect\citeauthoryear{Imada et al.}{2015}]{Imada15}  
 Imada, S., Murakami, I., \& Watanabe, T.: 
 2015, Physics of Plasma,  22, 101206
 \bibitem[\protect\citeauthoryear{Innes et al.}{2003}]{Innes03}  
 Innes, D.~E., et~al.: 
 2003, \solphys, 217, 267
  \bibitem[\protect\citeauthoryear{Kamio et al.}{2007}]{Kamio07}  
 Kamio, S., Hara, H., Watanabe, T., et al.:
 2007, \pasj , 59, S757
  \bibitem[\protect\citeauthoryear{Kamio et al.}{2010}]{Kamio10}  
 Kamio, S., Hara, H., Watanabe, T., Fredvik, T., \& Hansteen, V. H.:
 2010, \solphys , 266, 209
  \bibitem[\protect\citeauthoryear{Kosugi et al.}{2007}]{Kosugi07}  
 Kosugi, T., Matsuzaki, K., Sakao, T., et al.: 
 2007, \solphys , 243, 17
  \bibitem[\protect\citeauthoryear{Kumar et al.}{2019}]{Kumar19}  
 Kumar, P., Karpen, J. T., Antiochos, S. K., Wyper, P. F., DeVore, C. R., \& DeForest, C. E.: 
 2019, \apj, 873, 93
  \bibitem[\protect\citeauthoryear{Lemen et al.}{2012}]{Lemen12}  
 Lemen, J. R., Title, A. M., Akin, D. J., et al.: 
 2012, \solphys , 275, 40
  \bibitem[\protect\citeauthoryear{Liu et al.}{2011}]{Liu11}  
 Liu, C., Deng, N., Liu, R., et al.: 
 2011, \apjl , 735, L18
    \bibitem[\protect\citeauthoryear{Madjarska}{2011}]{Madjarska11}  
 Madjarska M. S.:
 2011, \aap, 526, A19
 \bibitem[\protect\citeauthoryear{Matsui et al.}{2012}]{Matsui12}  
 Matsui, Y., Yokoyama, T., Kitagawa, N., \& Imada, S.: 
 2012, \apj , 759, 15
 \bibitem[\protect\citeauthoryear{Milligan \& Dennis}{2009}]{Milligan09}  
 Milligan, R. O., \& Dennis, B. R.: 
 2009, \apj , 699, 968
 \bibitem[\protect\citeauthoryear{Miyagoshi \& Yokoyama}{2004}]{Miyagoshi04}  
 Miyagoshi, T., \& Yokoyama, T.: 
 2004, \apj , 614, 1042
 \bibitem[\protect\citeauthoryear{Moore et al.}{2010}]{Moore10}  
 Moore, R. L., Cirtain, J. W., Sterling, A. C., \& Falconer, D. A.: 
 2010, \apj , 720, 757
  \bibitem[\protect\citeauthoryear{Moore et al.}{2013}]{Moore13}  
 Moore, R. L., Sterling, A. C., Falconer, D. A., \$ Rome, D.: 
 2013, \apj , 769, 134
 \bibitem[\protect\citeauthoryear{Moreno-Insertis\& Galsgaard}{2013}]{Moreno13}  
 Moreno-Insertis, F., \& Galsgaard, K.: 
 2013, \apj , 771, 20
   \bibitem[\protect\citeauthoryear{Narukage et al.}{2014}]{Narukage14}  
 Narukage N., Sakao T., Kano R., et al.: 
 2014, \solphys , 289, 1029 
   \bibitem[\protect\citeauthoryear{Ogawara et al.}{1991}]{Ogawara91}  
 Ogawara, Y., Takano, T., Kato, T., et al.: 
 1991, \solphys , 136, 1 
   \bibitem[\protect\citeauthoryear{Panesar et al.}{2018}]{Panesar18}  
 Panesar, N. K., Sterling, A. C., \& Moore, R. L.: 
 2018, \apj, 853, 189 
  \bibitem[\protect\citeauthoryear{Shen et al.}{2017}]{Shen17}  
 Shen, Y., Liu, Y. D., Su, J., Qu, Z., \& Tian, Z.: 
 2017, \apj, 851, 67 
  \bibitem[\protect\citeauthoryear{Shibata et al.}{1992}]{Shibata92}  
 Shibata, K., Ishido, Y., Acton, L, W., et al.: 
 1992, \pasj , 44, L173
   \bibitem[\protect\citeauthoryear{Shibata et al.}{1994}]{Shibata94}  
 Shibata, K., Nitta, N., Matsumoto, R., et al.: 
 1994, in X-Ray Solar Physics from Yohkoh, ed. Y. Uchida, T. Watanabe, K. Shibata, \& H. S. Hudson (Tokyo: Universal Academy Press), 29
  \bibitem[\protect\citeauthoryear{Shimojo et al.}{2001}]{Shimojo01}  
 Shimojo, M., Shibata, K., Yokoyama, T., \& Hori, K.: 
 2001, \apj, 550, 1051 
 \bibitem[\protect\citeauthoryear{Sterling et al.}{2015}]{Sterling15}  
 Sterling, A. C., Moore, R. L., Falconer, D. A., \& Adams, M.: 
 2010, \nat, 523, 440
 \bibitem[\protect\citeauthoryear{Sterling et al.}{2017}]{Sterling17}  
 Sterling, A. C., Moore, R. L., Falconer, D. A., Panesar, N. K., \& Martinez, S.: 
 2017, \apj, 844, 28
  \bibitem[\protect\citeauthoryear{Strong et al.}{1992}]{Strong92}  
 Strong, K. T., Harvey, K., Hirayama, T., et al.: 
 1992, \pasj , 44, L161
   \bibitem[\protect\citeauthoryear{Tian et al.}{2012}]{Tian12}  
 Tian, H., McIntosh, S, W., Xia, L., et al.: 
 2012, \apj , 748, 21
   \bibitem[\protect\citeauthoryear{Tsuneta et al.}{1991}]{Tsuneta91}  
 Tsuneta, S., Acton, L., Bruner, M., et al.: 
 1991, \solphys , 136, 37
 \bibitem[\protect\citeauthoryear{Tsuneta et al.}{2008}]{Tsuneta08}  
 Tsuneta, S., Ichimoto, K., Katsukawa, Y., et al.: 
 2008, \solphys , 249, 167
  \bibitem[\protect\citeauthoryear{Wyper et al.}{2018}]{Wyper18}  
 Wyper, P. F., DeVore, C. R., \& Antiochos, S. K.: 
 2018, \apj, 852, 98
   \bibitem[\protect\citeauthoryear{Yokoyama \& Shibata}{1995}]{Yokoyama95}  
 Yokoyama, T., \& Shibata, K.: 
 1995, \nat , 375, 42
   \bibitem[\protect\citeauthoryear{Yokoyama \& Shibata}{1996}]{Yokoyama96}  
 Yokoyama, T., \& Shibata, K.: 
 1996, \pasj , 48, 353
  \bibitem[\protect\citeauthoryear{Young \& Muglach}{2014a}]{Young14a}  
Young, P. R., \& Muglach, K.: 
 2014, \pasj, 66, S12
  \bibitem[\protect\citeauthoryear{Young \& Muglach}{2014b}]{Young14b}  
 Young, P. R., \& Muglach, K.: 
 2014, \solphys , 289, 3313
  \bibitem[\protect\citeauthoryear{Young}{2015}]{Young15}  
 Young, P, R.: 
 2015, \apj , 801, 124
\end{thebibliography}
%
% Without BibTeX 

%%%%%%%%%%%%%%%%%%%%%%%%%%%%%%%%%%%%%%%%%%%%%%%%%%%%%%%%%%%%%
% FIGURE CAPTIONS
%

%%%%%%%%%%%%%%%%%%%%%%%%%%%%%%%%%%%%%%%%%%%%%%%%%%%%%%%%%%%%%%
% TABLES
%

\begin{table}
\begin{tabular}{c|ccccccc}
	\hline
	wavelength of AIA (\AA) & 131 & 335 & 193 & 304 & 94 & 171 & 211 \\
	\hline
	apparent velocity (km/s) & 77 & 71 & 64 & 61 & 37 & 55 & 57  \\
	\hline
\end{tabular}
\caption{Apparent velocities of the jet in each wavelength calculated from the observational results of AIA.}
\label{tab:aia}
\end{table}

\begin{table}
\begin{tabular}{c|cccc}
	\hline
	 & 184.54\AA~(Fe X) & 195.12\AA~(Fe XII) & 264.78\AA~(Fe XIV) & 284.16\AA~(Fe XV) \\
	\hline
	spire & -120 & -190 & -170 & -190 \\
	left leg & 180 & 230 & 190 & 50 \\
	right leg & 160 & 210 & 160 & 150 \\
	\hline
\end{tabular}
\caption{Doppler velocity (km/s) at each region and each wavelength measured by indicated line profile (Figure~\ref{fig:line}).}
\label{tab:vel}
\end{table}

\end{article} 
\end{document}